\documentclass{elsart}     % for users with LaTeX2e

\input epsf
 
\begin{document}
\begin{frontmatter}
\title{ Continuum model for radial interface growth}
 
\author[ANU]{M.T. Batchelor} 
\author[UNSW]{B.I. Henry}
 and
\author[UNSW]{S.D. Watt}
\address[ANU]{Department of Mathematics, School of Mathematical
Sciences, Australian National University, Canberra ACT 0200,
Australia}
\address[UNSW]{Department of Applied Mathematics, School of Mathematics,
University of New South Wales, Sydney NSW 2052, Australia}
 
\begin{abstract}
A stochastic partial differential equation along the lines of 
the Kardar-Parisi-Zhang equation is introduced for the evolution 
of a growing interface in a radial geometry.
Regular polygon solutions as well as radially symmetric solutions
are identified in the deterministic limit. The polygon solutions,
of relevance to on-lattice Eden growth from a seed in the zero-noise limit,
are unstable in the continuum in favour of the symmetric solutions.
The asymptotic surface width scaling for stochastic radial interface
growth is investigated through numerical simulations and found to be
characterized by the same scaling exponent as that for stochastic growth
on a substrate.
\end{abstract}
{\small Short title: {\it Radial interface growth}}
\end{frontmatter}

\section{Introduction}
The Kardar-Parisi-Zhang (KPZ) equation \cite{KPZ}
\begin{equation}
\frac{\partial h}{\partial t}=\nu\frac{\partial^2 h}{\partial x^2}
+\frac{\lambda}{2}\left(\frac{\partial h}{\partial x}\right)^2+\eta(x,t)
\label{kpzeq}
\end{equation}
is the simplest nonlinear
stochastic evolution equation for a growing interface.
In this equation $h(x,t)$ denotes the height of an  interface
at substrate position $x$ and time $t$;
$\eta(x,t)$ denotes a white noise term
and $\nu$ and $\lambda$ are growth parameters
related to surface tension and lateral growth respectively.
Among its applications the KPZ equation models vapor deposition at
large length scales, 
Eden growth on a substrate,
directed polymers, and one-dimensional turbulence.
In the case where $\lambda=0$ the KPZ equation reduces to the linear 
Edwards-Wilkinson (EW) equation \cite{EW} for  random deposition.
For reviews of the extensive literature on the 
KPZ, EW and related equations and their applicability to evolving 
interfaces in a wide range of physical phenomena, we refer the 
reader to Refs. \cite{FV,V,BS,M}.

In this paper we are motivated by one of the above applications, 
Eden growth, to obtain an analogous equation to the KPZ equation 
in the case when the growth is radial from a seed, rather than
from the usual substrate.
We derive this equation in Section 3, after recalling the
derivation of the KPZ equation in Section 2.
In Section 4 we peform linear stability analysis for deterministic
radially symmetric solutions. 
Numerical solutions are reported in Section 5, along with the
exponent $\beta$ governing the width of the interface.
We conclude with a summary.

\section{KPZ equation for growth on a substrate}
The KPZ equation comprises 
three terms;
surface tension, lateral growth
and noise.
The surface tension acts to smooth out the interface by rounding off
bumps and filling in hollows. It is assumed that the time rate of change
in the local height function due to surface tension is proportional
to the surface curvature
\begin{equation}
\kappa=
\frac{\frac{\partial^2h}{\partial x^2}}{\left[1+
(\frac{\partial h}{\partial x})^2\right]^{\frac{3}{2}}}.
\end{equation}
The basis of the lateral growth contribution in the KPZ equation
is the assumption that all points on the
growing interface move with uniform speed
$v$ in a direction normal to the interface.
The geometry underlying lateral growth in the KPZ equation is shown in fig. 1.
The lateral growth
contribution to the time rate of change of the height function
is found by projecting
the outwards normal growth in a direction orthogonal to the
substrate. Ignoring overhangs, the result which follows from
the Pythagorean theorem (fig. 1) is 
\begin{equation}
v\left[1+\left(\frac{\partial h}{\partial x}\right)^2\right]^{\frac{1}{2}}.
\end{equation}
The noise term in the KPZ equation is assumed to 
be uncorrelated white noise; 
\begin{eqnarray}
\langle \eta(x,t) \rangle&=&0 \label{noise1}\\
\langle \eta(x,t)\eta(x',t')\rangle&=&2D\delta(x-x')(t-t')\label{noise2}
\end{eqnarray}
where $D$ is a surface diffusion constant.

The reduction of 
\begin{equation}
\frac{\partial h}{\partial t}=
\nu \frac{\frac{\partial^2h}{\partial x^2}}{\left[1+
(\frac{\partial h}{\partial x})^2\right]^{\frac{3}{2}}}+
v\left[1+\left(\frac{\partial h}{\partial x}\right)^2\right]^{\frac{1}{2}}
+\eta(x,t)
\end{equation}
to the KPZ equation, Eq. (\ref{kpzeq}), is acheived by making a small
gradient approximation
$\frac{\partial h}{\partial x}\ll 1$
and transforming to a co-moving frame.

\section{Continuum equation for radial growth from a seed}
We now consider
surface tension, lateral growth and noise in a radial geometry.
In this case the interface is characterized by its radial
position $R(\theta,t)$ at angle $\theta$.
We assume that $R(\theta,t)$ is single valued so that there are
no `overhangs' along a direction of constant $\theta$.
The noise term is now a function of $\theta$ rather than $x$
but otherwise it is identical to Eqs. (\ref{noise1}), (\ref{noise2}).
The curvature of the surface characterized by $R(\theta,t)$
is given by
\begin{equation}
\frac{R^2+2\left(\frac{\partial R}{\partial \theta}\right)^2
-R\frac{\partial^2 R}{\partial \theta^2}}
{\left[R^2+\left(\frac{\partial R}{\partial 
\theta}\right)^2\right]^{\frac{3}{2}}}.\label{radcur}
\end{equation}

The geometry for 
lateral growth in a radial geometry is shown in fig. 2.
Simple trigonometric manipulations lead to the lateral growth term
\begin{equation}
\frac{v}{R}\left[R^2+\left(\frac{\partial R}{\partial 
\theta}\right)^2\right]^{\frac{1}{2}}.
\label{radlat}
\end{equation}
Combining the radial noise term with surface tension proportional
to the surface curvature, Eq. (\ref{radcur}), and the lateral
growth given by Eq. (\ref{radlat}) then gives
the continuum model for radial growth of an interface:
\begin{equation}
\frac{\partial R}{\partial t}=
\frac{v}{R}\left[R^2+\left(\frac{\partial R}{\partial 
\theta}\right)^2\right]^{\frac{1}{2}}
%-\nu \frac{\left[R^2+2\left(\frac{\partial R}{\partial \theta}\right)^2
-\nu \frac{R^2+2\left(\frac{\partial R}{\partial \theta}\right)^2
%-R\frac{\partial^2 R}{\partial \theta^2}\right]}
-R\frac{\partial^2 R}{\partial \theta^2}}
{\left[R^2+\left(\frac{\partial R}{\partial 
\theta}\right)^2\right]^{\frac{3}{2} }}
+\eta(\theta ,t).\label{radkpz}
\end{equation}

The deterministic version of Eq. (\ref{radkpz}) has
radially symmetric solutions
$R(\theta, t)=f(t)$ where
\begin{equation}
\frac{d f}{d t}=\frac{\nu}{f}+v.
\end{equation}
In the absence of surface tension the radially symmetric solutions 
have the form $R(t)=vt+C$ whereas with the inclusion of surface tension
$R(t)$ satisfies the transcendental equation
\begin{equation}
\nu\log (\nu+v R(t))-vR(t)+v^2t=C.
\end{equation}
Another class of deterministic
`solutions' are polygons defined by straight line segments
\begin{equation}
R(\theta, t)=\frac{v t}{\cos(\theta -c)}
\end{equation}
where the parameter $c$ is fixed between the vertices of the
polygon. These solutions have zero curvature
along the faces however they break down
at the vertices where they are not differentiable.
A special class are the regular  $n$ sided polygons
\begin{equation}
R_j(\theta, t)=\frac{v t}{\cos(\theta -\frac{\pi}{n}-\frac{2j\pi}{n})},\quad
\frac{2\pi j}{n}\le \theta\le
\frac{2\pi (j+1)}{n}
\label{poly}
\end{equation}
with $j=0,\ldots, n-1$.
We mention these `solutions' here because exact regular polygons 
are obtained in on-lattice zero-noise simulations of the Eden A model 
from a seed, where on the square lattice, $n=4$ and on the triangular 
and honeycomb lattices $n=6$.

\section{Linear stability analysis of deterministic solutions}

The linear stability of solutions $\bar R$
to the continuum model for radial growth can be investigated by substituting
the perturbed solution $R=\bar R+\rho$ into the growth equation,
Eq. (\ref{radkpz}), and retaining terms to first order in the perturbation
$\rho$. This results in a linear partial differential equation to
solve for the perturbation. In the special case where $\bar R =\bar R(t)$ 
are the radially symmetric solutions the perturbation satisfies
\begin{equation}
\frac{\partial \rho}{\partial t}=\frac{\nu}{\bar R(t)^2}\rho
+\frac{\nu}{\bar R(t)^2}\frac{\partial^2\rho}{\partial\theta^2}.
\end{equation}
Assuming a separable solution $\rho=T(t)H(\theta)$ where
 $H(\theta)=H(\theta+2\pi)$
now results in
\begin{equation}
\rho(\theta,t)=\frac{a_0}{2}e^{\nu\omega(t)}+\sum_{m=1}^\infty 
(a_m\cos(m\theta)+b_m\sin(m\theta))
e^{(1-m^2)\nu\omega(t)}\label{linsta}
\end{equation}
where 
\begin{equation}
\omega(t)=\int\frac{1}{\bar R^2(t)}dt
\end{equation}
is strictly positive.
The coefficients $a_m, b_m$ are determined by the initial
conditions $\rho(\theta,t_0)$. In particular,
\begin{eqnarray}
a_m&=&\frac{1}{\pi e^{(1-m^2)\nu\omega(t_0)}}\int_{-\pi}^{\pi} 
\rho(\theta,t_0)\cos(m\theta) d\theta.\\
b_m&=&\frac{1}{\pi e^{(1-m^2)\nu\omega(t_0)}}\int_{-\pi}^{\pi} 
\rho(\theta,t_0)\sin(m\theta) d\theta.
\end{eqnarray}
Hence if
\begin{equation}
\int_{-\pi}^{\pi}\rho(\theta,t_0)d \theta=
\int_{-\pi}^{\pi} \rho(\theta,t_0)\cos(\theta) d\theta=
\int_{-\pi}^{\pi} \rho(\theta,t_0)\sin(\theta) d\theta=0 \label{cc}
\end{equation}
then the perturbation will decay in time and the radially symmetric 
solutions will be asymptotically stable.

We can use the above analysis to
investigate the linear stability of the regular polygon solutions.
The idea is to consider an initial regular polygon profile
as an initially perturbed radially symmetric solution $\bar R(t)$. In this
case the perturbation initially satisfies 
\begin{equation}
\rho(t_0)=R_j(\theta, t_0)-\bar R(t_0)
\end{equation}
where $R_j(\theta, t_0)$ is defined in Eq. (\ref{poly}).
It is a simple exercise to verify that the conditions in Eq. (\ref{cc}) 
are met in this case and thus the starting regular polygon will
relax to the radially symmetric solution.
This is a direct consequence of surface tension suppressing growth
at the vertices. The persistence of regular polygon solutions in zero noise
Eden growth is due to a balance between this  surface tension and
the underlying lattice anisotropy. 

\section{Numerical investigations}

In general the stability of arbitrary profiles can be investigated directly
by numerically integrating the continuum equation
for radial interface growth with the given profile
as an initial condition.
In the numerical studies reported below we have employed the
simple discretizations
\begin{eqnarray}
R(\theta,t)&=&R(i\Delta\theta,j\Delta t)=R_{i,j}\\
\eta(\theta,t)&=&\eta(i\Delta\theta,j\Delta t)=\eta_{i,j}\\
\frac{\partial R}{\partial t}&=&\frac{R_{i,j+1}-R_{i,j}}{\Delta t}\\
\frac{\partial R}{\partial\theta}&=&
\frac{R_{i+1,j}-R_{i-1,j}}{2\Delta\theta}\\
\frac{\partial^2 R}{\partial\theta^2}&=&
\frac{R_{i+1,j}+R_{i-1,j}-2R_{i,j}}{\Delta\theta^2}.
\end{eqnarray}

Fig. 3 shows time snapshots (at intervals of 100 time units)
of the growing
interface starting from a diamond in the deterministic limit,
$\eta=0$, with growth velocity $v=0.1$
for the two cases; (a) $\nu=0$ (solid line) and (b) $\nu=0.1$
(dashed line). In this simulation
 $\Delta t=1/20$ and $\Delta\theta=2\pi/500$.
In each case the diamond profile is smoothed out and approaches
a circular profile in agreement with the predictions
of the linear stability analysis above.
In the integration with surface tension this
approach to the circular profile is faster.

Fig. 4 shows time (number of iterations) snapshots of the growing interface
starting from a circle in a stochastic integration
with $\eta$ a random number in the range
$[-0.008,0.008]$ and physical parameters $v=1.0,\nu=0.1$. In this simulation
 $\Delta t=1/1000$ and $\Delta\theta=2\pi/100$.
The growing interface in Fig. 4 is similar in
appearance to the growing interface in Eden growth \cite{BH}.

Fig. 5 shows a log-log plot of the surface width versus 
time (number of iterations), averaged over five stochastic integrations, 
with parameters as in Fig. 4. Also shown on this plot is a straight line 
of best fit with slope $\beta=1/3$. This suggests that the continuum 
equation, Eq. (\ref{radkpz}), for radial interface growth is in
the same universality class as the KPZ equation.

\section{Summary}

We have derived the stochastic partial differential equation,
Eq. (\ref{radkpz}), for the evolution of a growing interface in a 
radial geometry.
The equation has regular polygon solutions
as well as radially symmetric solutions in the 
deterministic limit.
Linear stability analysis reveals that the regular polygon solutions
are unstable in favour of the radially symmetric solutions.
The numerical solution of the fully stochastic equation
indicates that the width of the interface scales with the
same exponent, $\beta = 1/3$, as the KPZ equation.
Presumably this result can be obtained exactly and thus
provide explicit confirmation that both continuum models of
interface growth lie in the same universality class.

\ack
This work has been supported by the Australian Research Council.

\newpage

\begin{figure} 
\vspace{200mm}
\includegraphics{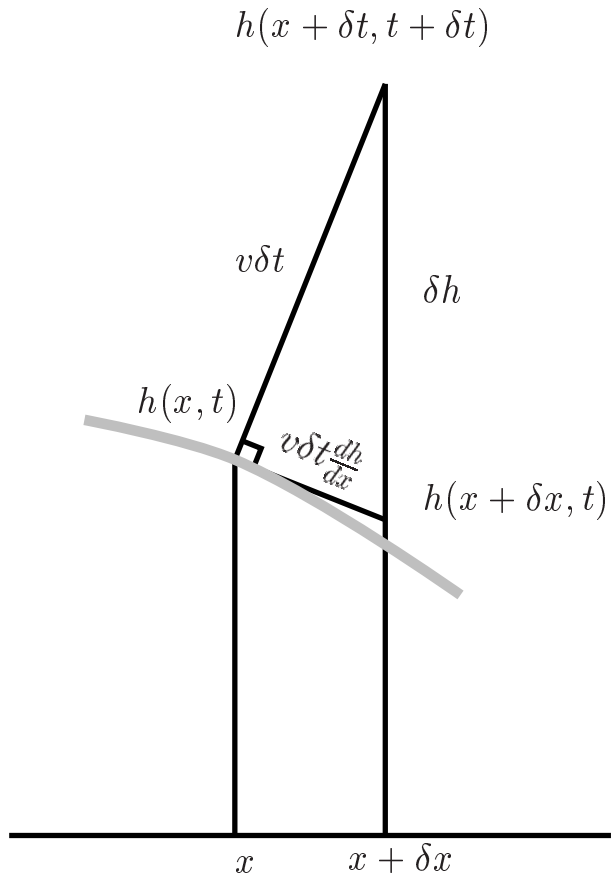}
\caption{
Geometry for interface growth from a substrate. The interface
is shown as the thick grey line. Here 
$(\delta h)^2 = (v \delta t)^2 \left[1 + \left( \frac{d h}{d x} 
\right)^2 \right]$.
} 
\label{fig1}
\end{figure}

\begin{figure}
\vspace{200mm}
\includegraphics{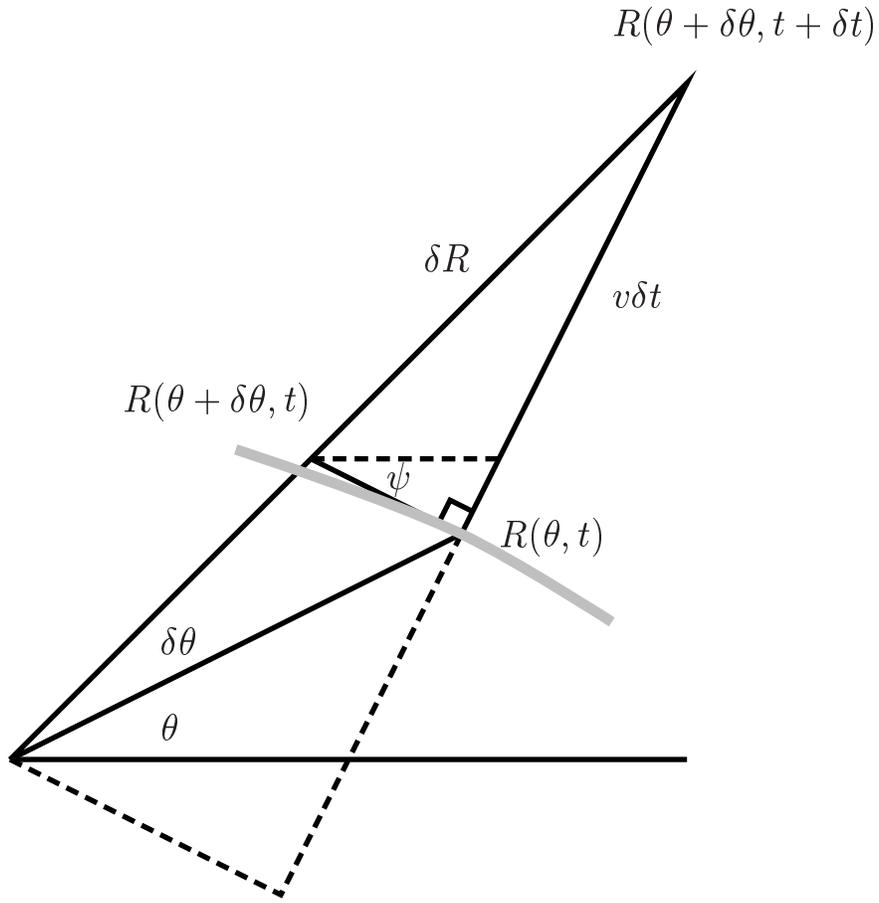}
\caption{
Geometry for radial interface growth from a seed.
The interface is shown as the thick grey line.
Here 
$(\delta R)^2 = \frac{(v \delta t)^2 \left[ R^2 + (v \delta t)^2 +
2 R v \delta t \sin(\theta + \psi) \right]}
{\left[R \sin(\theta + \psi) + v \delta t \right]^2}$.
}
\label{fig2}
\end{figure}

\begin{figure}
\vspace{200mm}
\includegraphics{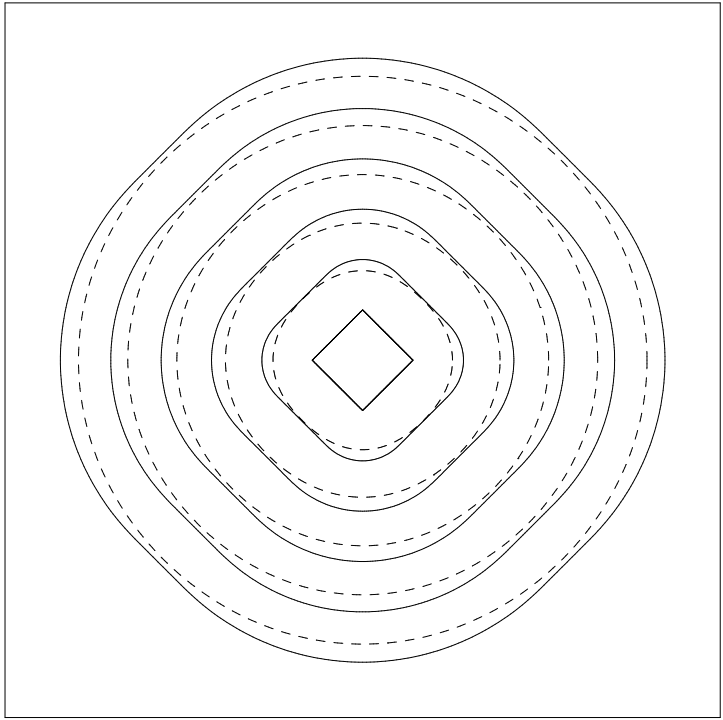}
\caption{Time snapshots of the evolution of an initial diamond shaped
interface in the deterministic limit of the continuum radial
growth equation. The solid line shows growth without surface tension
and the dashed line is growth with surface tension.}
\label{fig3}
\end{figure}

\begin{figure}
\vspace{200mm}
\includegraphics{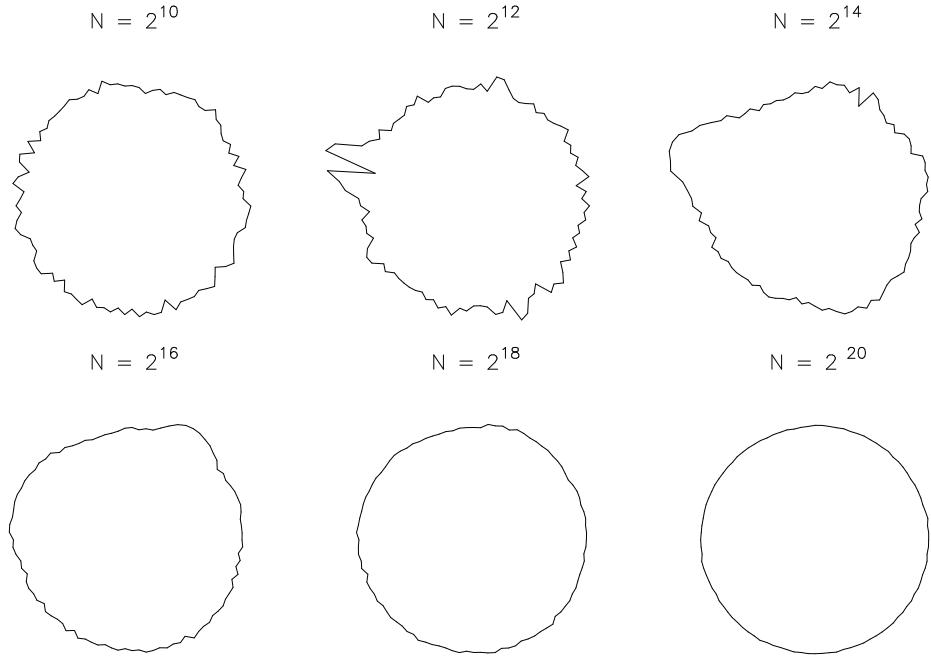}
\caption{Time snapshots of the evolution of an initial circular 
 interface
in the stochastic continuum radial
growth equation. $N$ is the number of iterations
with time step $1/1000$.}
\label{fig4}
\end{figure}

\begin{figure}
\vspace{200mm}
\includegraphics{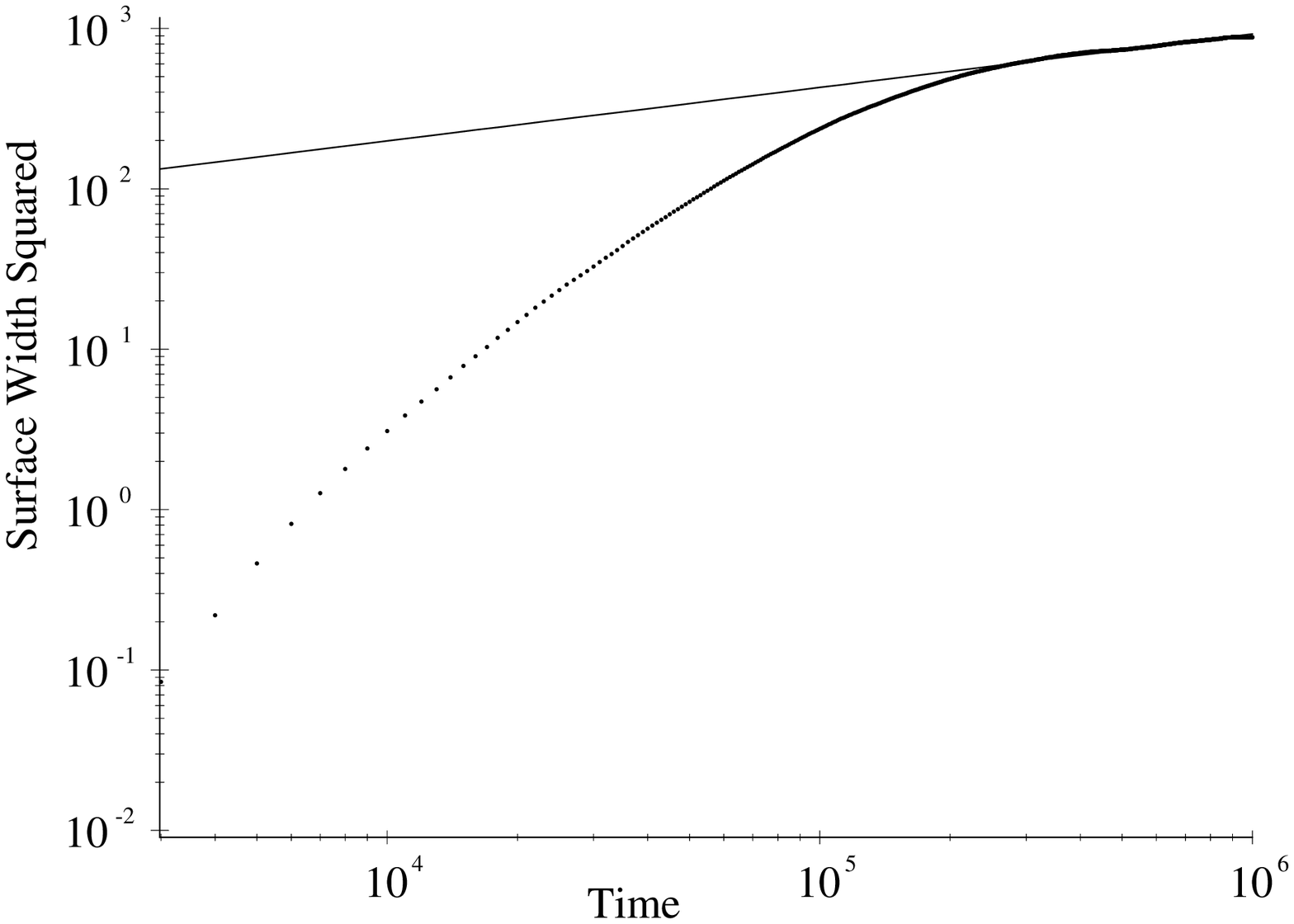}
\caption{Log-log plot of the surface width squared versus time
(number of iterations)
in the stochastic continuum radial
growth equation. The solid line is the best fit line
over the domain $t\in(2.5\times 10^5,10^6)$
with slope $\beta=1/3$.}
\label{fig5}
\end{figure}


\begin{thebibliography}{9}
\bibitem{KPZ} M. Kardar, G. Parisi and Y.-C. Zhang,
Phys. Rev. Lett. 56 (1986) 889.
\bibitem{EW} S.F. Edwards and D.R. Wilkinson,
Proc. Roy. Soc. Lond. A 381 (1982) 17.
\bibitem{FV} F. Family and T. Vicsek, Dynamics of Fractal Surfaces 
(World Scientific, Singapore, 1991).
\bibitem{V} T. Vicsek, Fractal Growth Phenomena, 2nd edition 
(World Scientific, Singapore, 1992).
\bibitem{BS} A.-L. Barab\'asi and H.E. Stanley, Fractal Concepts in
Surface Growth (Cambridge University Press, Cambridge, 1995). 
\bibitem{M} P. Meakin, Fractals, Scaling and Growth Far From
Equilibrium (Cambridge University Press, Cambridge, 1998)
\bibitem{BH} M.T. Batchelor and B.I. Henry, Phys. Lett. A 157 (1991) 229. 
\end{thebibliography}
\end{document}